# Single crystal monolithic up-converter solar cell device tandems with integrated optics

*Georgios E. Arnaoutakis[1*], Elena Favilla[2], Mauro Tonelli[2,3] and Bryce S. Richards[4,5*]*

[1] Department of Mechanical Engineering, Hellenic Mediterranean University, Estavromenos, Heraklion Crete 710 04, Greece

[2] Dipartimento di Fisica - Universita' di Pisa, Largo B. Pontecorvo, 3 I-56127 PISA, Italy

[3] Megamaterials srl Largo B. Pontecorvo , 3 56127 I Pisa

[4] Institute of Microstructure Technology (IMT), Karlsruhe Institute of Technology, Hermann-von-Helmholtz-Platz 1, 76344 Eggenstein-Leopoldshafen, Germany

[5] Light Technology Institute (LTI), Karlsruhe Institute of Technology, Engesserstrasse 13, 76131 Karlsruhe, Germany

*Corresponding author: E-mail: bryce.richards@kit.edu (Prof. Bryce Richards), arnaoutakis@hmu.gr (Dr Georgios Arnaoutakis)

**Abstract**
Solar photons possessing energy less than the band-gap of a single-junction solar cell can be utilized via the up-conversion (UC) of two or more photons, resulting in the emission of a single above-bandgap photon. Due to the non-linear nature of UC, highly concentrated light is required, which is typically much greater than the practical concentration limits of a solar cell. It has been proposed that concentrating up-conversion solar cells (UC-SC) with optical elements integrated into the device could help realize the high solar irradiance required. To avoid scattering problems arising from common UC materials based on micro-crystalline powders, in this work concentrators are investigated with mono-crystalline up-converters in silicon-based tandem devices. An external quantum efficiency (EQE) of 6% with 1493 nm infrared illumination at 876 W/m$^2$ was obtained in upconverter device with concave integrated optics. At an irradiance higher than 90 W/m$^2$ (equivalent to 2.95× in the 1450-1600 nm range), the non-concentrating UC-SC exhibited 1.5× higher EQE than the UC-SC with CPC, while below 90 W/m$^2$ the CPC UC-SC exhibited 1.95× higher EQE than the non-concentrating reference device. Due to the negligible scattering of the UC layer, the distribution of localized irradiance is revealed along with its effect on the performance of devices. It is found that irradiance is accumulated within the first 1 mm of the UC layer with peaks at variable depths according to the concentrating scheme. These results suggest ample space for improved up-conversion devices by using integrated optics.

## 1. Introduction

Up-conversion (UC) is one of the emerging third generation photovoltaic (PV) approaches to improve the sub-bandgap response of single-junction solar cells. It involves the conversion of two or more low energy photons into one photon possessing energy greater than the bandgap. For crystalline silicon (c-Si) solar cells with an (indirect) bandgap of $E_g$ = 1.12 eV, such near infrared (NIR) transparency losses account for more than 20% of the energy in the solar spectrum [1], and capturing this could improve the conversion efficiency of a solar cell by 20% [2,3]. To date, the majority of UC research applied to solar energy harvesting has been oriented towards c-Si solar cells as this technology represents >94% of today's PV production globally [4]. The vast majority of c-Si PV modules are designed to operate under one-sun illumination (1000 W/m$^2$) conditions. There exist some exceptions where c-Si solar cells operated at solar concentrations ranging from 10 suns (10,000 W/m$^2$) [5], to 40 suns [6] up to 92 suns [7,8]. Solar cells made from c-Si are not suited for operation at higher solar concentration due to series and contact resistance [9], but more important Auger recombination beginning to dominate and significantly reducing the conversion efficiency [10]. This latter fact is unfortunate as UC is a multi-photon non-linear process and thus the required solar concentration on the up-converter can be more than an order of magnitude higher [11–13]. A recent review by Richards *et al.* indicated that at least a 100-times increase in the incident photon flux (or alternatively the intermediate state lifetime of the UC mechanism) is required to start seeing efficient solar UC, noting that, even then, many assumptions were very optimistic [14]. Trivalent erbium ($Er^{3+}$)-doped up-converters are suitable for silicon solar cell tandems, due to their efficient up-conversion of ~1500 nm light to ~980 nm emission, which lies just above the bandgap of c-Si. Such up-converters exhibit an increased quantum yield at higher than 350 suns [12] with an onset of saturation at 100,000 suns [15], although noting the level of saturation is dependent on other factors such as $Er^{3+}$ ion concentration [16] and it has been recently reported that thermal effects definitely need to be considered at high concentration [17]. Broadband upconverters in the 1500-1750nm as well as the 1100-1400 nm band would be promising for a greater utilization of the solar spectrum. In addition to the approach of sunlight concentration presented in the current work, there are a number of routes for enhancing upconverters discussed in detail in [14]. A broader spectrum can be achieved by co-doping $Er^{3+}$ with additional lanthanide ions. $Dy^{3+}$, $Yb^{3+}$ and $Ho^{3+}$ have been reported as sensitizers. Transition metals such as $Ni^{2+}$ exhibit wide absorption in 1100-1400 nm. A high energy transfer efficiency to $Er^{3+}$ was recently reported in $Ni^{2+}$-, $Er^{3+}$-doped ceramics [18]. Quantum dots were also reported to shift the 1100-1400 nm band [19]. Both reports focused on energy transfer to the first excited level of erbium, despite several metastable levels exist in this band. Unfortunately, the limited quantification of the absolute upconversion quantum yield (PLQY) does not permit an evaluation of the practical gains for a solar cell. Two of the most efficient upconverters reported to-date are $BaY_2F_8$:30%$Er^{3+}$ with internal UCQY=10.1% and *β*-$NaYF_4$:25%$Er^{3+}$ with PLQY=12% at 0.4 W/cm². These upconverters resulted in 8% and 4.4% solar cell EQE at 0.4 W/cm² infrared illumination, respectively [20]. The EQE of organic upconverters approaches 30% as summarized by Pedrini and Monguzzi [21]. However, the conversion is limited in the 340-1010 nm range, therefore not yet suitable for silicon solar cells.

In a previous work, the present authors demonstrated that a higher concentration of light on the up-converter can be achieved via the inclusion of optical elements between the bifacial c-Si solar cell and the up-converter [22]. Furthermore, an efficiency enhancement equal to the concentration ratio of 2.8× has been reported by using compound parabolic concentrators (CPC) [23]. This use of optics enabled the optimization of the solar concentration independently from the overlying solar cell [24,25]. Such results were demonstrated using a micro-crystalline powder of hexagonal erbium-doped sodium yttrium fluoride (*β*-NaYF: 25%$Er^{3+}$), one of the best performing up-converters for c-Si solar cells to-date [20]. For an extensive recent review of up-conversion materials, solar cells and devices, the interested reader is directed towards reference [20] as well as the more recent critical review of UC for PV and photocatalysis, based on photophysical analysis [21]. It is important to clarify that another micro-crystalline up-converter, gadolinium oxysulfide ($Gd_2O_2S$: 10%$Er^{3+}$) exhibited higher quantum yield under monochromatic excitation. However, under broadband excitation, more relevant to the application in solar cells, *β*-$NaYF_4$:25%$Er^{3+}$ exhibited a higher quantum yield [26], and further demonstrated external quantum efficiency (EQE) of 1.79% at 1000 W/m² [27] and a $\Delta$Jsc of 9.4 mA/cm² due to the up-converter at 94 suns [28] with c-Si solar cells.

In this work, we seek to move away from the use of micro-crystalline powders. With a crystal size of 2-200 μm [13], increased scattering inhibits the knowledge of light distribution inside the micro-crystalline up-converters and its performance. A previous work by Arnaoutakis *et al.* [23] found an increase in EQE from 1.33% to 1.80% by the integration of CPC. In this paper, motivated by the high quantum yield, EQE and negligible scattering of erbium-doped barium yttrium fluoride ($BaY_2F_8$:30%$Er^{3+}$) single crystals [29,30], we investigate the effect of the concentrators on mono-crystalline up-converters and their performance on up-converter solar cell tandems.

## 2. Materials and Methods

The different configurations investigated within this work are shown in **Figure 1(a-d).** Reference devices (a) comprised of an up-converter layer (red) and a planar, bifacial c-Si solar cell (blue) were investigated together with different optical elements: (b) a concave reflector, (c) a CPC of acceptance angle 25°, and (d) a CPC of acceptance angle 45°. In (e) the main transitions leading to conversion of near 1500 nm light in the up-converter, are displayed in the energy level diagram. Excitation around 1493 nm will be absorbed via ground state absorption (GSA) to the $^4I_{13/2}$ level. From this level, energy can be transferred and up-converted (ETU) resonantly between two nearby $Er^{3+}$ ions. Depending on the population of each level, process ETU1 between $^4I_{13/2}$ - $^4I_{9/2}$, ETU2 between $^4I_{11/2}$ - $^4F_{9/2}$ or ETU3 between $^4F_{9/2}$ - $^4F_{7/2}$ may occur. From the upper levels of each process, following several multi-phonon relaxations, emission to the ground state can occur resulting to light around 517 nm, 650 nm, 790 nm and 970 nm [30,31].

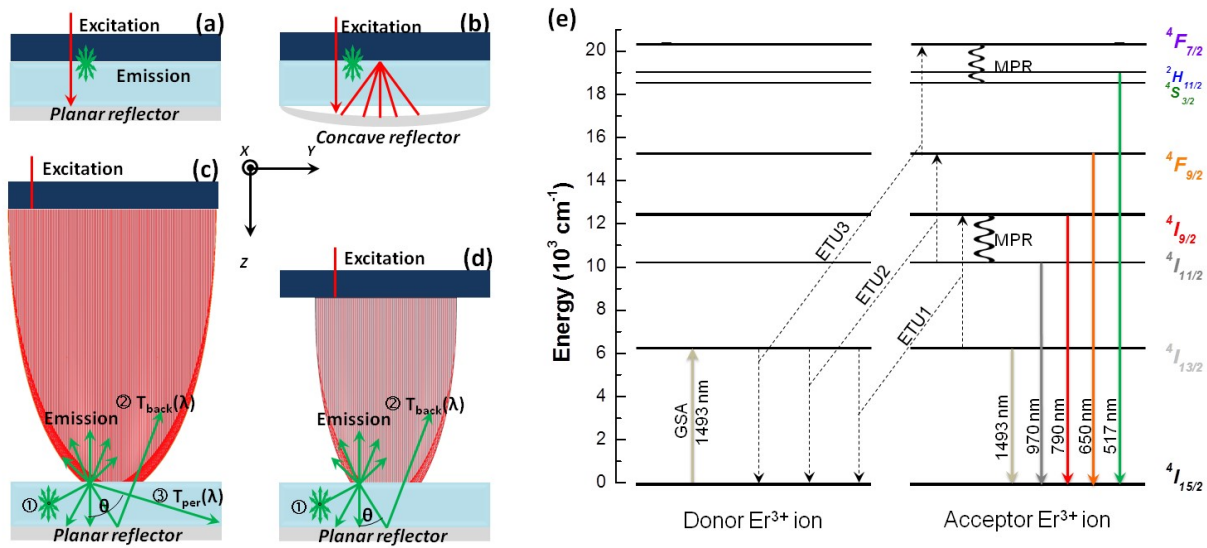

Figure 1: Four different configurations are investigated in this paper, with the up-converter shown in ruby and the bifacial c-Si solar cell drawn in blue. In the reference scenarios, the excitation is at normal incidence and the emission is reflected by a specular planar back reflector. Note that no concentrating element is employed in the reference scenario. In (b) the concave reflector concentrates any non absorbed excitation at the up-converter solar cell interface. In (c) and (d), the introduction of the CPC-25 and CPC-45 elements respectively into the configuration results in NIR light entering the up-converter at oblique angles. The isotropic up-converted emission is externally coupled back into the CPC and delivered to the overlying bifacial c-Si solar cell. Internally, the emission may be ① re-absorbed and re-emitted, ② reflected at the back reflector and transmitted towards the solar cell, or ③ lost via the peripheral facets of the up-converter. (d) The energy level diagram depicts the main transitions taking place in the $Er^{3+}$-doped up-converter with upward arrows indicating the absorption of light, downward arrows the resulting emission, (--) non-radiative energy transfer up-conversion (ETU) and (~) multi-phonon relaxation (MPR).

## 2.1. Simulation

The geometrical loss mechanisms occurring in the studied configurations were calculated using ray optics and Monte Carlo simulations (Optis, Optisworks) in three-dimensional models using >$10^7$ rays over an area of 100 $mm^2$. Non-sequential ray tracing was used with 0.5 mm facet width used for tesselation of each optical element. The simulation included the refractive indices and absorption coefficients of the CPC optics (Schott B270) and the $BaY_2F_8$:30%$Er^{3+}$ material itself [31,32]. Empirical data [33] were used for the refractive index of Schott B270, while for $BaY_2F_8$ coefficients were fit in a modified Sellmeier equation of the form [32]:

$$n^2 = A + \frac{B\lambda^2}{\lambda^2 - C} + \frac{D\lambda^2}{\lambda^2 - E}$$

, where λ is the wavelength and the Sellmeier coefficients A=1.04037, B=1.12565, C=0.00686, D=0.2038 and E=66.043. The resultant fit is in good agreement with the dispersion of $BaY_2F_8$:30%$Er^{3+}$ recently determined by high accuracy variable angle spectroscopic ellipsometry [34]. Further details on the simulation can be sourced in [35].

Near-infrared (NIR) excitation light at the peak absorption wavelength (1493 nm) and a Gaussian intensity profile was used with power density, $P_{exc}$, normally incident on the top surface of the devices (as depicted in **Figure 1(a-d)**. Planar wave illumination accurately represents non-concentrated air-mass 1.5 global (AM1.5G) sunlight, while Gaussian illumination better matches concentrated sunlight conditions. The rationale for using Gaussian illumination in the simulations was primarily driven by the assumption that a low concentration optical element is more likely to be used for primary concentration on top of silicon to reduce the concentration requirement of the secondary integrated optical element. Gaussian illumination also describes the distribution of lasers used for excitation, thereby more accurately simulating the conditions used in the presented experiments.

The NIR excitation was reduced by the transmittance of the up-conversion optimized c-Si solar cell, that is 85% at 1493 nm, while the c-Si solar cell exhibits a rear-side reflectance of 8% for the UC light emitted between 980-1050 nm [36]. The emitted power, $P_{em}$, was calculated as $P_{em}(\lambda) = P_{exc}(\lambda) \times iPLQY(\lambda)$ by considering the excitation power $P_{exc}$, the internal photoluminescence quantum yield (iPLQY) of the UC material at the main radiative UC emission from state $^4I_{11/2}$ to the ground state at 970 nm. It is important to note that the iPLQY is defined as the ratio of emitted photons to the number of absorbed photons, in contrast to the external PLQY (ePLQY) which is defined as the ratio of emitted photons to the number of incident photons. Thus, the ePLQY is more relevant for real-world devices as the impact of optical losses, such as incomplete absorption, are included in this figure-of-merit. The back reflectors were modeled as ideal specular. The wavelength dependent absorptance and path-length of reference and CPC cases were analytically calculated following the Beer-Lambert law as:

$$A_{Ref} = 1 - e^{-\alpha 2z}, A_{con} = 1 - e^{-\alpha 2z/\cos\theta}$$

, where z=0.233 cm, $\theta_{conc}$=50°, $\theta_{CPC-25}$=67°, $\theta_{CPC-45}$=35°.

## 2.2. Experiment

Specular planar reflectors based on thin films of gold (Au) of at at least 100 nm thickness were evaporated onto fused silica ($SiO_2$) slides. A silver-coated concave mirror with f = 17.5 mm, R = 35 mm and average infrared reflectance higher than 95% was used in a 2f configuration replacing the planar back reflector. Two geometries of CPC fabricated from Schott B270 glass with acceptance angles 25° and 45° (Edmund Optics, USA) were utilised in the experiments. The dimensions for these were: entry aperture ⌀9.01 mm and ⌀5.39 mm, and length 19.93 mm and 7.52 mm, respectively, while the exit aperture for both was ⌀2.5 mm. A tuneable laser (HP-Agilent, 8168-F) with a wavelength range of 1450-1590 nm, a typical output power of 6±1.4 mW, FWHM of 4 nm (spectral resolution-limited), and a collimated beam width of 4.2 mm was employed for EQE measurements. The excitation power density was characterized using a calibrated germanium photodiode (Newport, 818-IR) and a NIR camera (Electrophysics, Micronviewer 7290A), while the photogenerated current was measured with a source-meter (Keithley Instruments, 2440-C). More experimental details on the experimental setup can be found in a previous work [23]. The solar concentration $C_{exc}$ was calculated as [23]:

$$C_{exc} = \frac{\int P_{exc}(\lambda) \partial \lambda}{\int P_{AM1.5D}(\lambda) \partial \lambda} \quad (1)$$

, where $P_{exc}$ is the excitation power density or irradiance at the respective wavelength, $P_{AM1.5D}$ the spectral irradiance of the air-mass 1.5 direct (AM1.5D) solar spectrum [37] between $\lambda_1$ and $\lambda_2$, the integration limits corresponding to the absorption band of the up-converter.

The up-converter solar cells were based on planar bifacial solar cells fabricated on c-Si wafers, with a double-layer anti-reflection coating (ARC) – comprised of 110 nm hydrogenated silicon nitride (a-SiN$_x$:H) and 110 nm magnesium fluoride (MgF$_2$) – on the front and a single-layer ARC of 120 nm a-SiN$_x$:H on the rear. The infrared absorptivity of silicon solar cells is highly sensitive to the texture angle. Riverola et.al. reported the dramatic reduction of λ=1-6 μm absorptivity of silicon when the texture angle is reduced from 60º to 0º [38]. These results were confirmed in the results of Rudiger et.al [36] comparing the transmittance and reflectance of planar and textured solar cells. In a planar solar cell, the infrared transmittance increased more than 20% compared to the textured cell, as shown in **Figure 2**. As a result, trapping of infrared photons in the cell was reduced. Without this trapping, however, reflection increased in the visible, reducing the EQE of the planar cell between 300-1200 nm as shown below.

Therefore, there is a trade-off between increasing infrared photons for the upconverter, while reducing visible photons and finally the photo-current of the solar cell. In a practical solar cell with texture, a lower infrared transmission will result to reduced upconverted light. At 1 sun, the ΔJsc from the sacrificial move from texturized to planar solar cell is 42.86-34.15=8.71 mA/cm². However, the record ΔJsc due to state-of-the-art upconverters was characterized at 94 suns. At this solar concentration the current density by the integral of the AM1.5D spectrum results to 357 mA/cm². A device based on the planar solar cells with solar concentration atop of the solar cell should produce at least a ΔJsc of 357-42=315 mA/cm² in the infrared to break even the sacrifices in the visible. At 94 suns the benefit from the upconverter is more than one order of magnitude lower than the sacrifice of the texture-to-planar solar cell. Consequently, combinations of different technologies in addition to solar concentration are required to enhance the performance of upconverters [14].

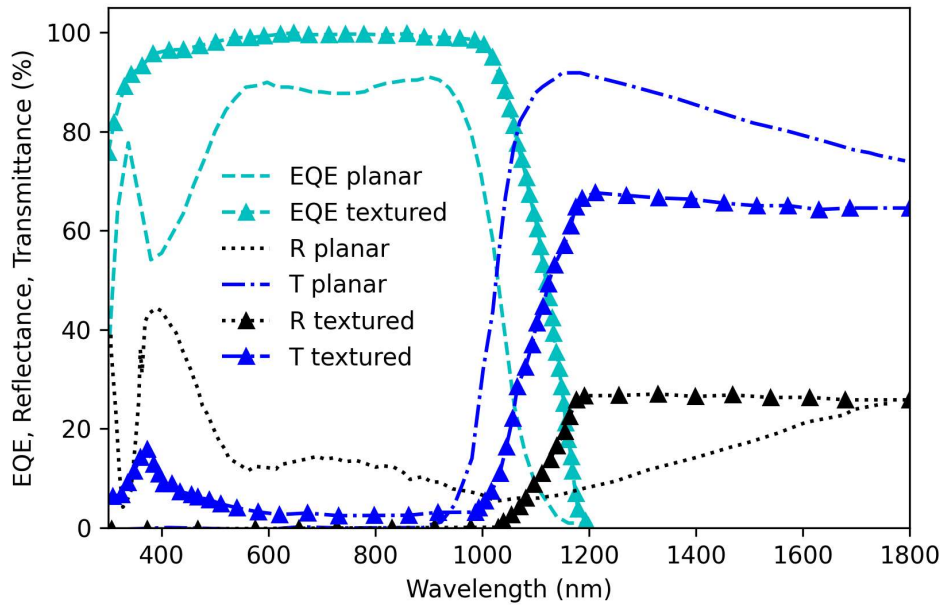

Figure 2: EQE, reflectance and transmittance of the planar bifacial solar cells utilized in the devices. Reflectance and transmittance was measured before metallization. Adapted from [36,39]

The absorptivity of silicon at λ = 1-6 μm is proportional to the concentration of the emitter and the BSF from $5 \times 10^{17}$ to $5 \times 10^{20}$ cm$^{-3}$ [38]. At a sheet resistance (Rsh) of 62 Ω/sq, a maximum phosphorus concentration of $3.3 \times 10^{20}$ cm$^{-3}$ can increase the absorption of a textured silicon wafer by approximately 40% [40]. An intuitive development towards reduced absorptivity of free carriers was reported in [36] via localized BSF centered near the highly reflecting contacts. Despite the low Rsh=19 Ω/sq of the BSF, the transmittance of the cells, did not show significant improvements. A boron-doped emitter profile with high sheet resistance of Rsh-130 Ω/sq was utilized, which may be associated with a high concentration of free carriers at the front side of the cells further limiting the infrared transmission. For further details on the optimization of the solar cells the reader is directed towards the work by [36].

$BaY_2F_8$:30%$Er^{3+}$ crystals were prepared under high purity argon conditions via the Czochralski process. Fluoride powders of 99.999% purity were melted at 972° C in a custom-built furnace, evacuated to $10^{-7}$ mbar to minimize contamination [31]. Finally, the samples were cut in rectangular blocks and polished to optical quality. The dimensions of the polished up-converter utilized here was 10.08 mm × 10.03 mm × 2.33 mm. The absorption coefficients were measured in an ultraviolet-visible-NIR spectrophotometer (Perkin-Elmer, Lambda 950).

## 3. Results and Discussion
### 3.1. Localized irradiance in the up-converter

The integrated optics concentrate light and locally change the distribution of the irradiance at the surface of the up-converter to improve the performance of the solar cell device [23]. In addition to the irradiance distribution at the surface of the up-converter, it is important to know the distribution of light inside a mono-crystalline up-converter and how this may affect its performance. The simulated local irradiance at 1493 nm at 210 W/m$^2$ as a function of depth $Z$ in the up-converter can be seen in **Figure 3** for the three cases illustrated in **Figure 1(a-c)**. The absorption coefficient of the up-converter (plotted in **Figure 5**) was considered in this simulation. The irradiance of the reference exhibits a log-normal distribution with the peak at 30 μm. The irradiance of CPC-25° and CPC-45°, in addition to the same log-normal distribution, exhibit additional peaks at $Z = 370$ μm and $Z = 720$ μm, proportional to the focal lengths of the CPC. A fraction of the light is transmitted without concentration by the CPC, resulting in a log-normal distribution. Excitation light reflected at the side walls of the parabola, see **Figure 1(c-d)**, will be focused inside the up-converter depending on its focal length. The focal length of the CPC-45° resides 720 μm inside the up-converter, while the absorption coefficient of $BaY_2F_8$: 30%$Er^{3+}$ at 1493 nm is 44 cm$^{-1}$. Therefore an amount of 1493 nm excitation will be absorbed before the focus. This dilute, compared to the laser beam, excitation, will be up-converted to visible wavelengths, as shown in the photographs of **Figure 3**. Due to this dilution and hence lower excitation irradiance to the reference beam, process ETU2 with emission at 650 nm may be more probable than ETU3 with emission at 550 nm, as qualitatively evidenced in the photograph. The irradiance profile of the concave reflector (not shown) is also estimated to have log-normal distribution with intensity proportional to $e^{-a_2 z/\cos\theta}$.

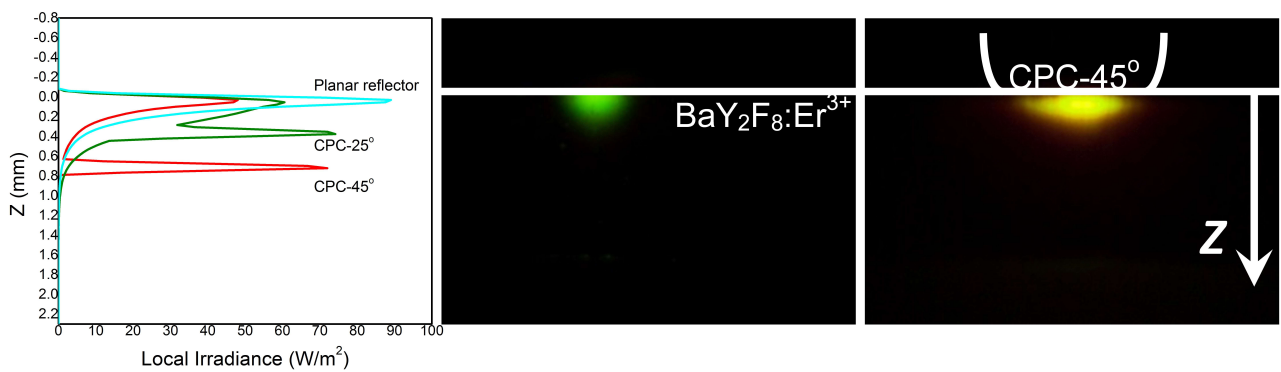

Figure 3: Simulated irradiance profile in the up-converter along the Z axis at the centre of the excitation beam. The CPC: up-converter interface is at 0 mm. Photographs of the reference (left) and CPC-45° (right) illuminated by 1493 nm light are also displayed inline.

The secondary peaks of CPC-25° and CPC-45° exhibit similar maximum irradiance, lower than that of the reference. The localized excitation peaks, correspond to centers with higher irradiance and consequently iPLQY. This result agrees with a higher iPLQY and ePLQY being measured for a thinner $BaY_2F_8$:30%$Er^{3+}$ sample [29], whereby a 1.76 mm-thick sample exhibited a higher probability of up-converted light being externally emitted. This is largely because a thinner sample suffers reduced re-absorption losses. For the investigated $BaY_2F_8$:30%$Er^{3+}$ up-converter, the re-absorption losses have been estimated to be as high as 51% [29].

The irradiance of emission at 970 nm at top of the up-converter interface that is directed to the solar cell (see Figure 1a-c) was simulated and shown in **Figure 4**. The emission is assumed to be a result of the excitation at 1493 nm at top of the up-converter. Therefore, it was simulated with the same illumination conditions, direct or concentrated (reference or CPC) illumination.

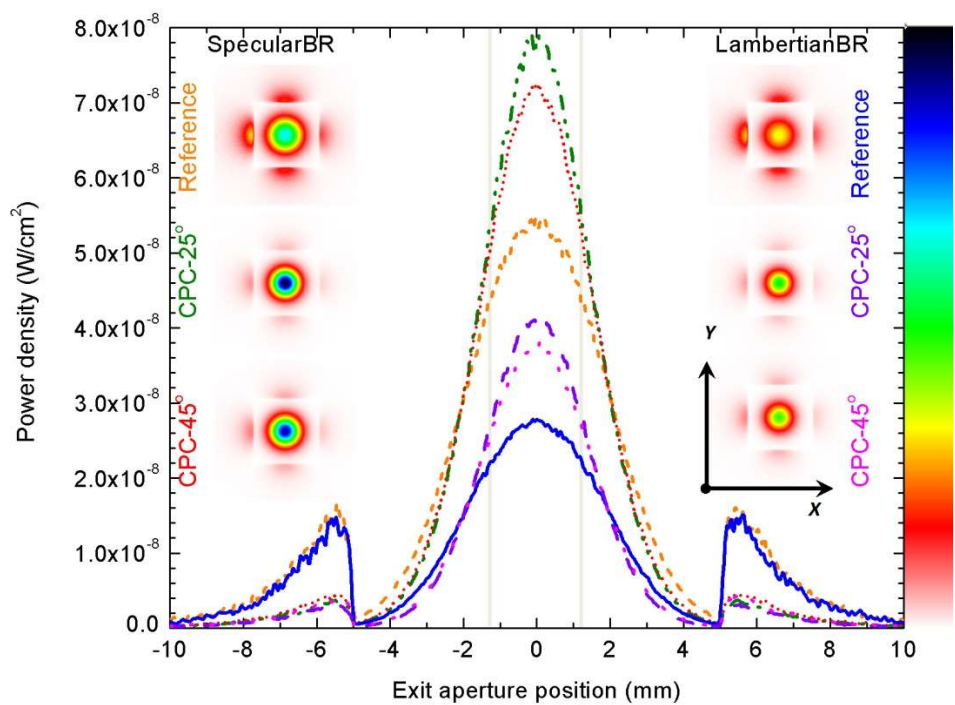

Figure 4: Irradiance profiles of the emission at 970 nm at the CPC : up-converter interface after reflection, simulated by ray-tracing. Six configurations are displayed for the reference (··· and —), CPC-25° (··−·· and — —) and CPC-45° (– – and ‑ ‑) with specular and Lambertian reflectors, respectively. The insets expose the spatial distribution of the irradiance transmitted out of the up-converter, with corresponding colours on the right.

The emission exhibits an irradiance profile consisting of one central peak as a result of the excitation beam and two lateral peaks outside the 10×10 mm up-converter (±5 mm of aperture position in X axis). The lateral peaks are higher for the reference than both CPC as a result of a wider illuminated area, 4.2 mm compared to 2.5 mm. Consequently, more 970 nm emission laterally escapes the up-converter in the reference compared to the CPCs. The intensity is higher for the specular back reflector than the Lambertian in all cases. This is a result of the Lambertian inducing a longer path-length in the up-converter and therefore higher absorption in the up-converter.

The emission can also be affected by the back reflector whether specular or Lambertian, however, for highly absorbing up-converters the contribution of this effect to EQE is expected to be negligible. Consequently, light emitted near the top of the up-converter and collected by the overlying solar cell should lead to a higher EQE which is presented in the next section.

### 3.2. External quantum efficiency - wavelength dependence

**Figure 5** displays the EQE of the investigated configurations illuminated at a constant power density of 90 W/m$^2$ or 2.95× concentrated atop the devices, the maximum available with our laser across the measured spectrum.

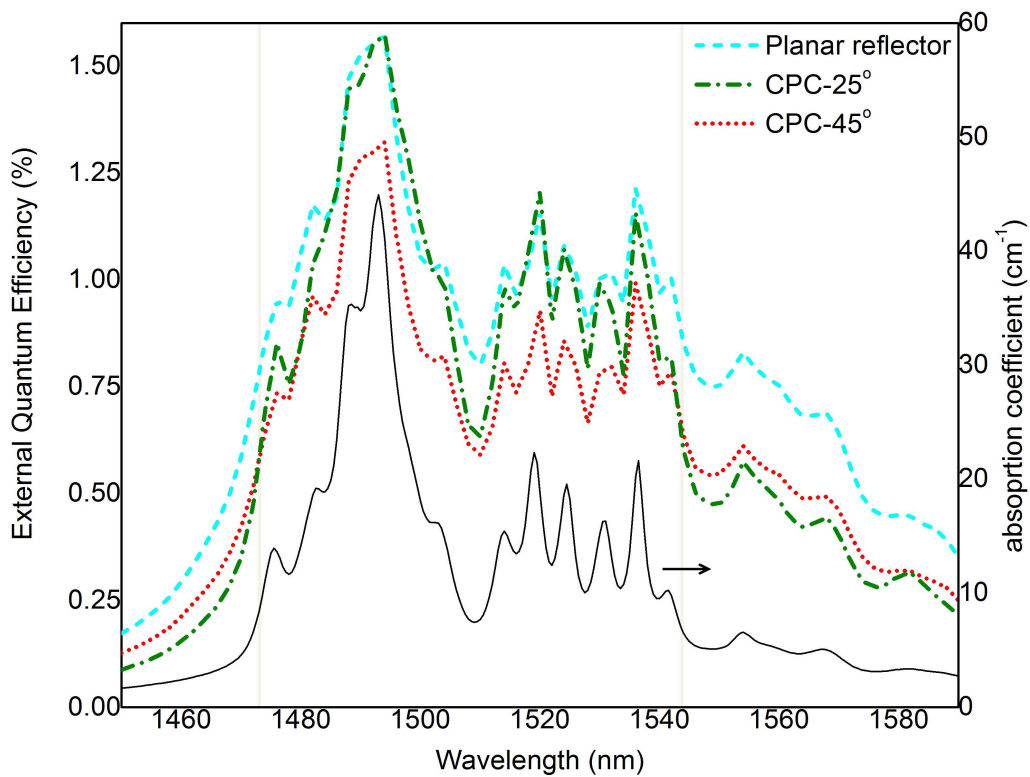

Figure 5: Measured EQE spectra of the UC-SC measured at 90 W/m$^2$ with: a) reference, b) CPC-45° c) CPC-25° geometries. The absorption tails of the up-converter are indicated by gray vertical lines.

The shape of all three EQE spectra closely match the absorption coefficient of BaY$_2$F$_8$: 30%Er$^{3+}$, reproducing the Stark structure of the $^4I_{15/2}$ - $^4I_{13/2}$ levels. This structure exhibits evidence of energy transfer up-conversion (ETU) as the dominant UC mechanism [41] common in highly Er$^{3+}$-doped materials [42]. The concentration of Er$^{3+}$ is 30% which promotes ETU and precludes excited state absorption from being the dominant population mechanism. The EQE of case a) reference and c) CPC-25° is higher than b) CPC-45° between 1472-1542 nm. Only in the absorption tails region (λ < 1472 nm and λ > 1542 nm, indicated by gray vertical lines) is the EQE using the CPC-45° fractionally higher than CPC-25°.

The EQE is the ratio of photo-generated carriers to the incident NIR photons. The obtained spectra can be explained in relation to the wavelength-dependent infrared absorptance and the path-length of the up-converter, displayed in **Figure 6**. At λ < 1472 nm and λ > 1542 nm, the low absorptance leads to different path-lengths in each case. CPC-25° induces the longest path-length, which is higher than CPC-45° and the reference. CPC-45° converges excitation light at an angle $\theta = 35.3°$ (see **Figure 1d**). At this angle, excitation light travels on average 1 cm longer path-length than the reference. However, in CPC-25° with an angle $\theta = 66.85°$ (see **Figure 1b**) excitation light has two-times longer path-length in the up-converter.

At the band 1472-1542 nm, however, the absorptance is >90% and the path-length is equal in all cases. It can be seen that the EQE of each configuration is proportional to the depth Z from the CPC: crystal interface. In the devices with planar back reflectors, the closer these high intensity regions are to the solar cell, the higher the collection and therefore the EQE will be. Upconversion is emitted spontaneously in a 4π solid angle. Depending on the path-length, several re-absorption events will reduce the upconversion externally emitted from the sample. Consequently, the high intensity excitations occurring near the surface and close to the solar cell, will propagate a shorter pathlength before being absorbed by the solar cell, resulting to a higher EQE. Consequently, the performance of mono-crystalline up-converter solar cells can be higher when the excitation is accumulated near the top surface of the up-converter. The absorptance of the concave mirror is also >90% at 1472-1542 nm, while lower than the CPC-25 and higher than the CPC-45 and reference outside this band. Contrary to the CPC, the focus of the concave mirror can be adjusted for higher performance near the top surface of the up-converter and is presented in the next section.

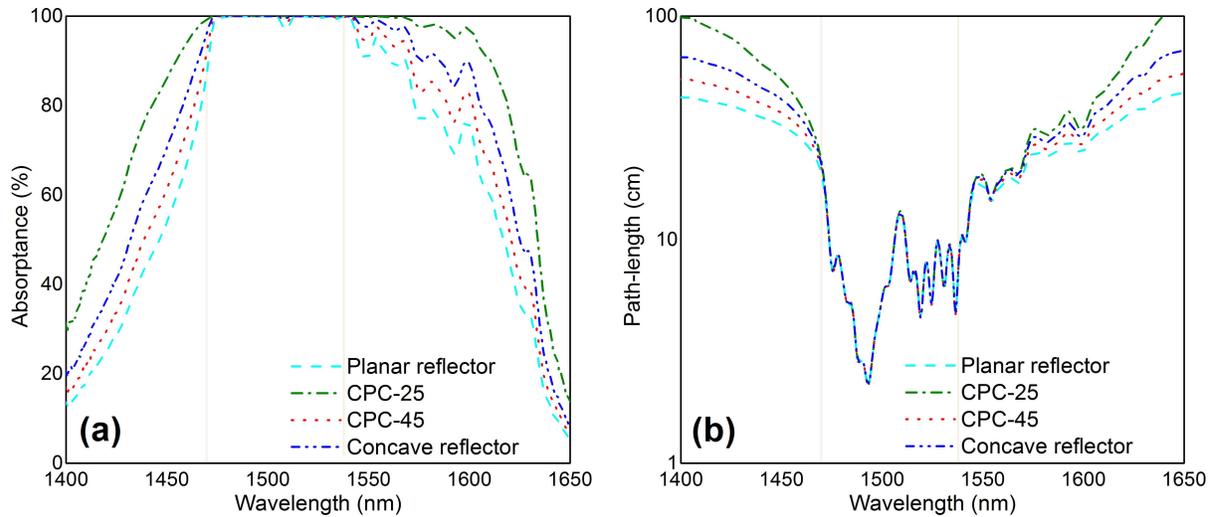

Figure 6: Absorptance (a) and path-length (b) of reference and CPC cases as a function of wavelength. The absorption tails of the up-converter are indicated by gray vertical lines.

### 3.3. Quantum efficiency - power dependence

The quantum efficiency was further characterized at the peak wavelength of 1493 nm as a function of power density and solar concentration. **Figure 7(a)** displays the EQE as a function of the power density incident on the device, while in **Figure 7(b)** the internal quantum efficiency (IQE) is displayed incorporating the transmittance through the c-Si solar cell and the concentrators.

At a low power density, the EQE of CPC-25° is higher than CPC-45°, planar and concave reflector devices. At power density higher than 90 W/m², the planar device performs better than CPC-25° and

CPC-45°. This is likely due to peripheral emission losses in the CPC devices, see **Figure 4**. As shown in **Figure 6(a)**, absorptance at 1493 nm is higher than 99% in all devices and aligns with a large fraction of absorbed pump power [43]. In the planar device, the absorbed power is concentrated at a shorter path-length in the up-converter as shown in **Figure 3**. The performance of the concave reflector device is lower than the CPC devices below 340 W/m² but higher above this power density. Excitation in the concave reflector device undergoes twice the path-length of the CPC devices. A longer path-length increases the re-absorption in the upconverter. As a result, the excitation intensity reaching the reflector is reduced, similar to emission at visible wavelengths in the photographs of Figure 3. Consequently, below 340 W/m2 the measured EQE results from upconverted light near the surface, while above this power density a higher external upconversion quantum yield (ePLQY) [12] results to more upconversion reaching the solar cell. Up-converted light as well as excitation is concentrated back to the up-converter by the concave reflector and finally absorbed by the solar cell. After taking into account the transmittance of the solar cell and the concentrators, the IQE of the devices is obtained with the highest IQE of 7.2% at 876 W/m² in the device with a concave reflector. The power conversion efficiency of the bifacial silicon cells increased by 0.19% under infrared excitation. The silicon cells without upconverter had an efficiency of 16.7% under AM1.5G conditions [27].

Several device effects are incorporated in the quantum efficiency in addition to a photoluminescence intensity measurement. These effects are: i) the spectral response of the solar cell absorbing up-conversion from all energy levels despite that higher than 90% is from the $^4I_{11/2}$ level, ii) the transmission of the solar cell at 1400-1600 nm, iii) a double path-length through the up-converter, and iv) the angular extent of the concentrators leading to variable local irradiance in the up-converter as presented in section 3.1 and further peripheral losses. The losses around the edges of the up-converter can be minimized by peripheral reflectors, as implemented with aluminized mylar [31]. It was demonstrated here that better performance obtains with concave reflectors in place of planar back reflectors. The used Ag-coated concave reflector has reduced reflectance around 1000 nm, thereby a higher IQE can be expected by using Au-coated reflectors.

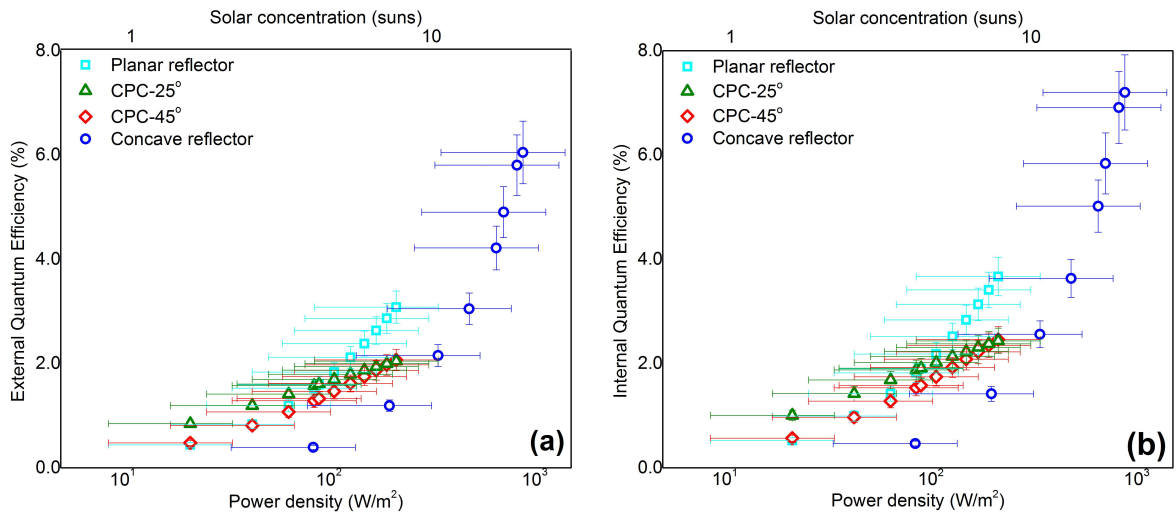

Figure 7: (a) EQE and (b) IQE of the up-conversion solar cell devices excited at 1493 nm as a function of incident power density atop the devices and equivalent solar concentration on semi-logarithmic scales.

The results suggest there is space for better performance from devices implementing higher concentration reflectors with mono-crystalline up-converters. Practical up-scaling of the integrated optics is promising given the practical advances in bifacial solar panels. To permit wider application, 1-sun operation atop of the devices is accommodating. Consequently, a concentration factor at least in the order of 30x, as evidenced in this report, can be accommodated by advances in micro-optics with ability for wide area application. CPC of acceptance angle less than 25° and 100 μm exit aperture [23] with shorter focal length will result in localized irradiance accumulated near the up-converter and a higher collection probability by the solar cell. A low concentration of 2.8× was attained with the presented concentrators. Higher solar concentration under sunlight is achieved in parabolic aplanat systems with secondary concentration. In these experiments, upconverters were characterized under solar concentration of 2300 suns [44]. In addition to increased quantum yield at higher concentration, a broader excitation can populate additional transitions in upconverters with potential for higher quantum yields [45].

Imaging optics would be able to concentrate to a focal spot near the diffraction limit, however, require high throughput and improved collection efficiency [22,46]. Optics developed for biomedical applications and bio-sensing with simultaneous capabilities of concentration and efficient collection can also be suitable to this end [24]. Specifically, dielectric microbeads [25] and microlens arrays [47,48] have demonstrated several orders of magnitude enhanced photonic devices, whereas double layer metasurfaces [49] exhibited enhancements by tailored confinement at the excitation and emission wavelengths.

## 5. Conclusions

The performance of mono-crystalline up-converter solar cell device tandems with integrated optics was presented in this paper. Solar cell devices coupled to a CPC with a lower acceptance angle (CPC-25°) exhibit higher IQE than the planar and CPC-45° below 1493 nm excitation at 90 W/m$^2$. At higher power density the CPC devices exhibit lower performance affected by the focal length of the CPC, which is shorter for the CPC-25° compared to the CPC-45°. In the CPC the local irradiance is accumulated 370 μm - 720 μm deeper in the up-converter, while in the planar device the irradiance is distributed near the top surface of the up-converter which is optically coupled to the solar cell, peaking at 30 μm. The highest IQE of 7.2% was measured with a concave reflector at 876 W/m$^2$ concentrating transmitted excitation and emission from the up-converter back to the solar cell. By knowing the irradiance distribution in the mono-crystalline up-converter and how this affects the performance of up-conversion solar cells, further improvements can be achieved by microspheric tailored optics as recently reported [25,47] and optimized up-converters with reduced re-absorption.

**Acknowledgements**

This research received the financial support of the Engineering and Physical Sciences Research Council (EPSRC) No. EP/I013245/1, the European Community's Seventh Framework Program (FP7/2007-2013) No. [246200] and the European Cooperation of Scientific & Technical Research (COST) via action CM1403. The authors wish to thank Jan C. Goldschmidt and Stefan Fischer from Fraunhofer Institute for Solar Energy Systems (Freiburg, Germany) for providing the silicon solar cell used in this work and the anonymous reviewers for providing valuable and constructive comments considerably improving this work. B.S.R. would like to acknowledge research funding from the Helmholtz Association: i) a Recruitment Initiative fellowship; and ii) the Research Field Energy – Program Materials and Technologies for the Energy Transition – Topic 1 Photovoltaics.